\documentclass[12pt,dvips]{article}
\usepackage[dvips]{graphicx}
\usepackage{graphicx}

\graphicspath{{plots/}}
\DeclareGraphicsExtensions{.eps.gz,.eps,.ps,.ps.gz}

\parskip=\itemsep               %?
\newcommand{\lwig}{\mbox{\,\raisebox{.3ex}
    {$<$}$\!\!\!\!\!$\raisebox{-.9ex}{$\sim$}\,}}
\newcommand{\gwig}{\mbox{\,\raisebox{.3ex}
    {$>$}$\!\!\!\!\!$\raisebox{-.9ex}{$\sim$}}\,}
\newcommand{\lambdabar}{{\hbox{$\lambda_e$\kern-1.9ex\raise+0.45ex\hbox{--}
\kern+0.2ex}}}

\newif\ifhepph

\hepphtrue

\ifhepph\date{\empty}\fi

\title{
\ifhepph{\normalsize%\rightline{Draft; not for distribution!}
\rightline{DESY 03-076}\rightline{hep-ph/0307034}}\fi
\vskip 1cm 
\bf\boldmath
An upper bound on the total cross-section for electroweak baryon %and lepton 
number violation
       \vspace{21mm}} 
\author{
A. Ringwald\\[4mm] 
Deutsches Elektronen-Synchrotron DESY, Hamburg, Germany}

\begin{document}
\begin{titlepage} 
  \maketitle
% declarations for front matter
%\vspace{1.cm}
\begin{abstract}
An upper bound on the total cross-section of s-wave electroweak 
instanton/sphaleron induced baryon plus lepton number violating 
processes is presented. It is obtained by exploiting a recently 
reported lower bound on the corresponding tunneling suppression 
exponent and by estimating the pre-exponential factor.
We find that the present knowledge about electroweak baryon plus lepton 
number violating processes still allows their eventual observability at 
the Very Large Hadron Collider, even as pure s-wave scattering. 
A possibly observable rate at cosmic ray facilities and neutrino telescopes 
is presently not excluded, but requires a substantial contribution from higher partial waves. 
\end{abstract}


\thispagestyle{empty}
\end{titlepage}
\newpage \setcounter{page}{2}

%\section{Introduction}
{\bf 1.}
Baryon and lepton number are not strictly conserved in the standard
electroweak model~\cite{'tHooft:up%,'tHooft:fv
}. In the background of topological non-trivial fluctuations of 
$SU(2)$ gauge fields  with topological charge $Q\not= 0$, notably of 
instantons ($Q= 1$) and anti-instantons ($Q= -1$)~\cite{Belavin:fg},  
baryon ($B$) and lepton ($L$) number change according to 
$\triangle B =\triangle L = -n_{\rm gen}\,Q$, where
$n_{\rm gen}=3$ is the number of fermion generations. Such fluctuations 
correspond to transitions between degenerate, topologically inequivalent
vacua~\cite{Jackiw:1976pf%,Callan:je
}. The latter are known to be separated by energy barriers whose 
minimum height is given by the static energy $E_{\rm sp}$ 
of an unstable static solution of the classical Yang-Mills equations called the 
sphaleron~\cite{Klinkhamer:1984di}. Its value\footnote{Here,  
$m_W = 80.423(39)$~GeV is the W$^\pm$ boson mass and $\alpha_W (m_W) = 0.033819(23)$ 
is the $SU(2)$ fine structure constant~\cite{Hagiwara:fs}.},  
$E_{\rm sp}\approx \pi\,m_W/\alpha_W\simeq 7.5$~TeV, 
sets the scale for non-perturbative $B+L$ violation in the Standard Model. 
Indeed, while these processes are exponentially suppressed at 
energies or temperatures below the sphaleron energy  by a tunneling factor, they 
are known to have a sizeable rate for 
temperatures above $E_{\rm sp}$~\cite{Kuzmin:1985mm%,Arnold:1987mh,Ringwald:1987ej
} and to have a crucial impact on the evolution of the baryon asymmetry of the universe. 
On the other hand, the long-standing question, first raised in 
Refs.~\cite{Aoyama:1986ej,Ringwald:1989ee%,Espinosa:qn
}, whether these processes occur with sizeable rates in high energy particle collisions, 
is still not finally settled (for reviews, see 
Ref.~\cite{Mattis:1991bj%,Tinyakov:1992dr,Guida:qy,Rubakov:1996vz
}).  

It is the purpose of this Letter to elaborate and to compare recent 
results on this issue~\cite{Khoze:1991mx,Ringwald:2002sw,Bezrukov:2003er%,Bezrukov:2003qm
} and to work out an upper bound on the total cross-section for 
electroweak baryon number violating processes. We will show that 
the present knowledge about the latter 
still allows their eventual observability at future 
colliders~\cite{Ringwald:2002sw,Farrar:1990vb%,Ringwald:1990qz,Gibbs:1994cw,Gibbs:1995bt
}, such as the Very Large Hadron Collider (VLHC)~\cite{vlhc}.  
We will also discuss implications for searches at cosmic ray facilities 
and neutrino telescopes~\cite{Morris:1991bb,%Morris:1993wg
Fodor:2003bn}.

\begin{figure}
\vspace{-2.5cm}
\begin{center}
\includegraphics*[width=12cm,clip=]{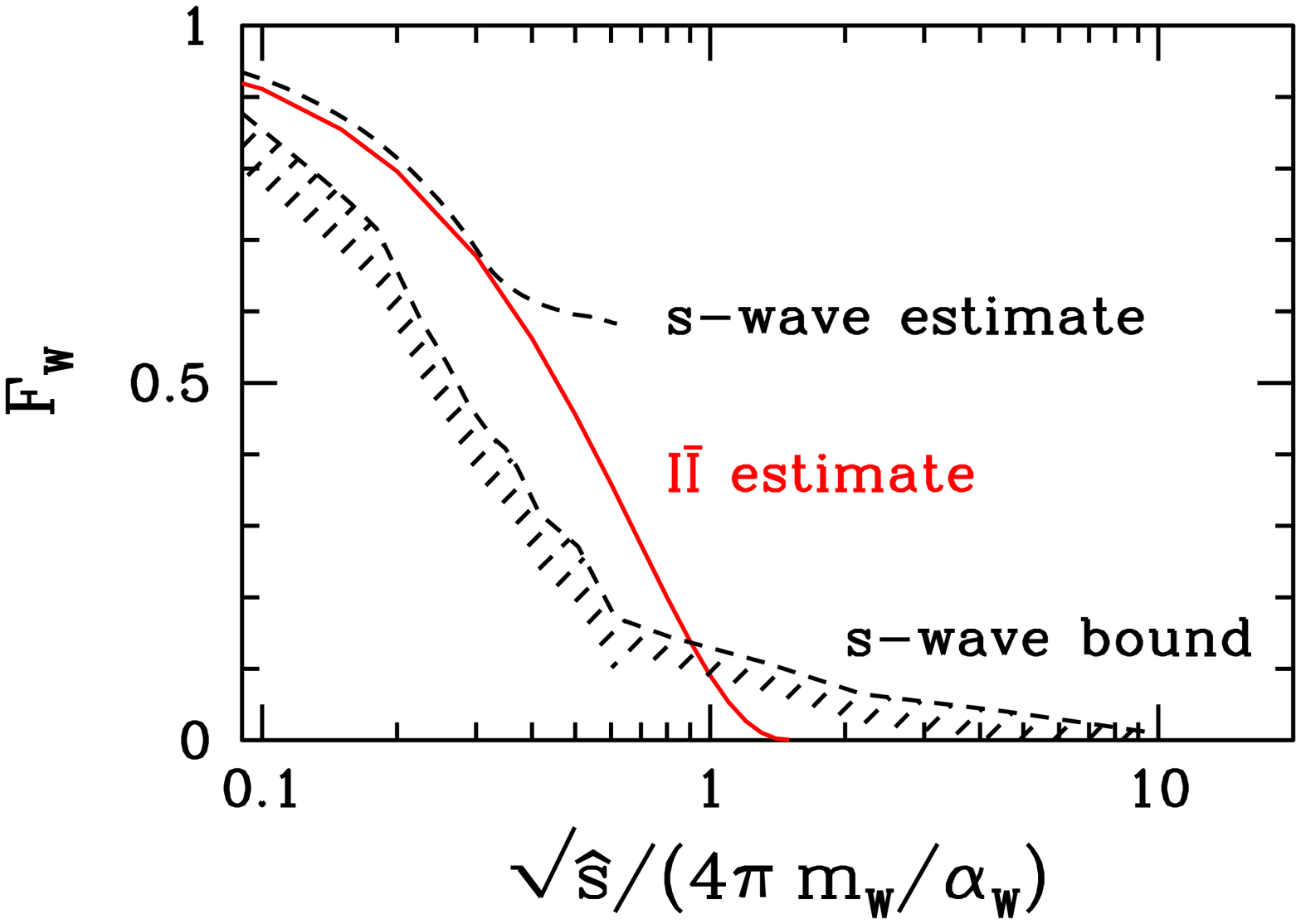}
\mbox{}\vspace{-3.5cm}
\includegraphics*[width=12cm,clip=]{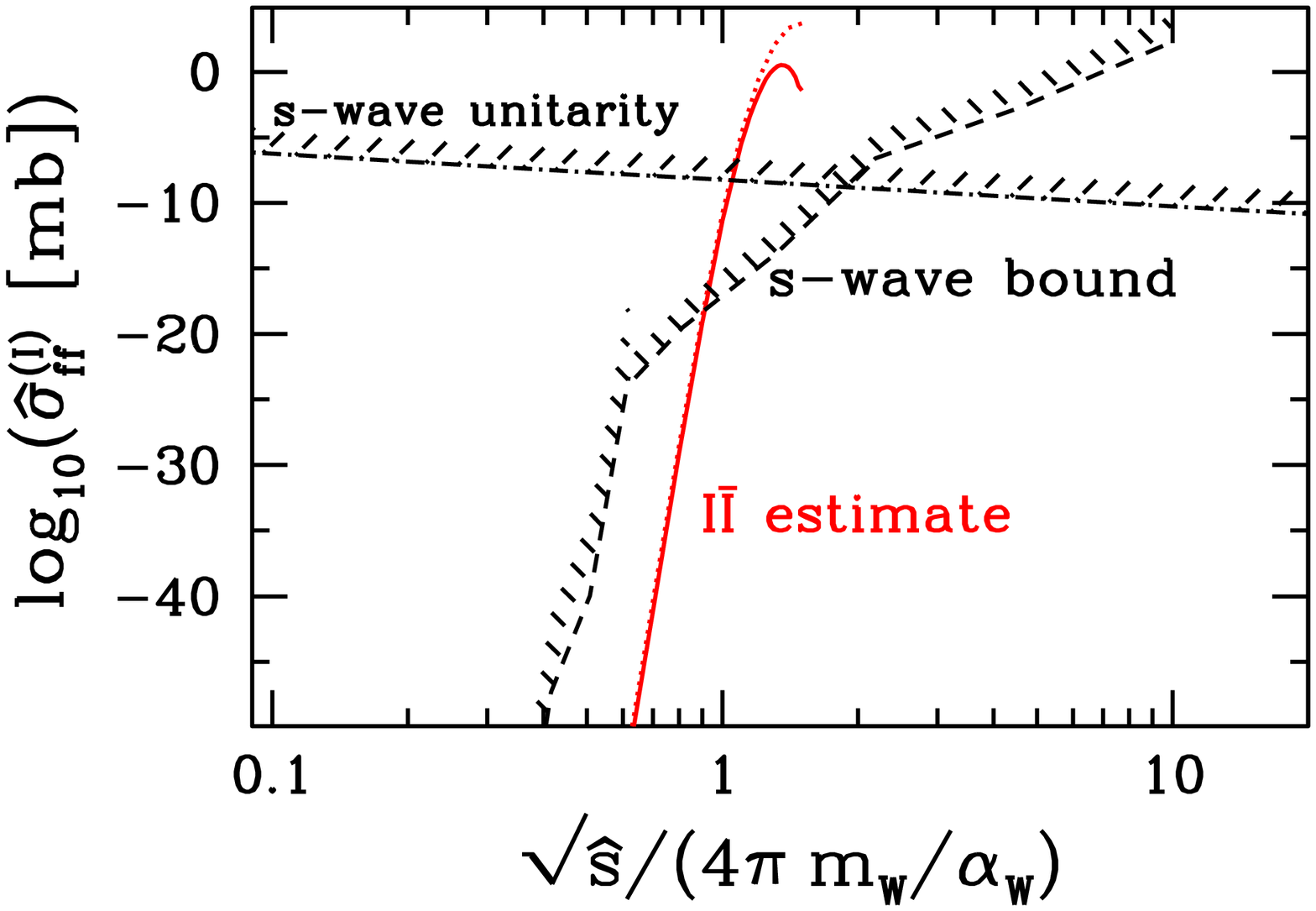}
\vspace{-1.5cm}
\caption[dum]{
{\em Top:} The tunneling suppression exponent $F_W$ as estimated from the  
$I\overline{I}$ interaction~\cite{Khoze:1991mx} (solid). Also shown is its estimate 
(dashed) and its lower bound (dashed shaded) 
for s-wave scattering~\cite{Bezrukov:2003er%,Bezrukov:2003qm
}. 
{\em Bottom:} The total cross-section $\hat\sigma^{(I)}_{\rm ff}$ 
for electroweak instanton-induced fermion-fermion scattering, ${\rm f+f}\stackrel{I}{\to}{\rm all}$, 
as estimated from the $I\overline{I}$ interaction, with complete calculation of the 
pre-exponential factors~\cite{Ringwald:2002sw} (solid) and with exploitation of the
estimate in Eq.~(\ref{cross-qfd}) for the pre-exponential factor (dotted). Also shown is 
the estimate (dashed) and upper bound (dashed shaded) on the total cross-section for s-wave scattering 
obtained from Eq.~(\ref{cross-qfd}) with exploitation of the estimate and the lower bound 
on the tunneling suppression exponent for s-wave scattering from Ref.~\cite{Bezrukov:2003er%,Bezrukov:2003qm
}, respectively. The s-wave unitarity bound (\ref{unit-lim}) is displayed also  
(dashed-dotted shaded).         
\label{holy-grail}}
\end{center}
\end{figure}

%\section{Setting the stage}
{\bf 2.}
At center-of-mass (CM) energies much less than the sphaleron energy, the rates
of anomalous electroweak $B+L$ violation are rapidly 
growing~\cite{Ringwald:1989ee%,Espinosa:qn
}.
The corresponding total cross-section is known to have an exponential 
form~\cite{McLerran:1989ab%,Khlebnikov:1990ue,Yaffe:1990iy,Arnold:1990va
}. Including essential pre-exponential factors~\cite{Khoze:1990bm}, 
one has, for the phenomenologically interesting case of 
fermion-fermion scattering via electroweak instantons/sphalerons, 
${\rm f+f}\stackrel{I}{\to}{\rm all}$,
\begin{eqnarray}
\nonumber
\hat\sigma_{\rm ff}^{(I)} 
&\approx & \frac{1}{m_W^2}
\,
\left( \frac{2\pi}{\alpha_W}\right)^{7/2}
\,
\exp\left[ -\frac{4\pi}{\alpha_W}\,
F_W \left( \frac{\sqrt{\hat s}}{4\pi m_W/\alpha_W} \right)\right]
\\[1.5ex] \label{cross-qfd} & \simeq &
5.3\times 10^3\ {\rm mb}\  
\exp\left[ -\frac{4\pi}{\alpha_W}\,
F_W \left( \frac{\sqrt{\hat s}}{4\pi m_W/\alpha_W} \right)\right]
\,,
\end{eqnarray}
where $\sqrt{\hat s}$ denotes the fermion-fermion CM energy.
By means of perturbative calculations of the relevant exclusive amplitudes 
about the instanton ($I$), squaring them and summing over the final states, or, alternatively, 
by means of a perturbative calculation of the forward elastic scattering amplitude 
about the widely separated instanton anti-instanton  ($I\overline{I}$) pair and 
determining the imaginary part to get the total cross-section via the optical theorem, one
may calculate the decisive tunneling suppression exponent $F_W$, which is sometimes
called ``holy-grail function''~\cite{Mattis:1991bj%,Tinyakov:1992dr,Guida:qy,Rubakov:1996vz
}, as a series in fractional powers of 
$\epsilon \equiv \sqrt{\hat s}/(4\pi m_W/\alpha_W)
\simeq \sqrt{\hat s}/(30\ {\rm TeV})$~\cite{Khoze:1990bm,Arnold:cx%,Diakonov:len,Mueller:fa,Diakonov:ur,Balitsky:xc
},
\begin{equation}
\label{FW-pert}
F_W (\epsilon ) = 1 - \frac{3^{4/3}}{2}\,  \epsilon^{4/3} + \frac{3}{2}\,\epsilon^2 + 
{\mathcal O}(\epsilon^{8/3})
\,.
\end{equation}
Correspondingly, the total cross-section (\ref{cross-qfd}) is exponentially growing
at $\epsilon\ll 1$. At $\epsilon = {\mathcal O}(1)$, however, the perturbative expression (\ref{FW-pert})
ceases to be helpful. In this energy regime, only 
estimates/extrapolations of 
and lower bounds on the tunneling suppression exponent exist (cf. Fig.~\ref{holy-grail} (top)), which
have been quantified recently~\cite{Khoze:1991mx,Ringwald:2002sw,Bezrukov:2003er%,Bezrukov:2003qm
}.

%%%%%%%%%%%%%%%%%%%%%%%%%%%%%%%%FIGURE%%%%%%%%%%%%%%%%%%%%%%%%%%
%%%%%%%%%%%%%%%%%%%%%%%%%%%%%%%%%%%%%%%%%%%%%%%%%%%%%%%%%%%%%%%%%

%\section{Estimates}
{\bf 3.}
In almost all theoretical investigations of electroweak $B+L$ violation in
high energy collisions, the pre-exponential factor in Eq.~(\ref{cross-qfd}) is not considered.
It is, however, numerically very large and 
has therefore to be taken into account before any conclusion on the observability 
of the effect can be drawn. Its size is mainly determined by the large universal factor 
$(2\pi/\alpha_W)^{7/2}\simeq 8.7\times 10^7$, where the exponent $7/2$ can be easily
understood as being the result of subtracting from the familiar nominal power 
$4\,N_c=8$~\cite{'tHooft:up%,'tHooft:fv
} the power $9/2$ -- nine being the effective number of collective coordinate saddle-point 
integrations~\cite{Khoze:1990bm,Ringwald:2002sw}. 
The large pre-exponential factor in Eq.~(\ref{cross-qfd}) leads to the fact that a cross-section 
$\hat\sigma^{(I)}_{\rm ff}\gwig 10^{-3}$~fb, observable
at the projected VLHC\footnote{The VLHC has a projected 
proton-proton CM energy of $\sqrt{s}=200$~TeV and a luminosity of about 
${\mathcal L}\approx 6\cdot 10^2$~fb$^{-1}$\,yr$^{-1}$~\cite{vlhc}.}~\cite{vlhc}, requires only 
a moderate reduction in the tunneling suppression exponent, $F_W\lwig 0.12$, from
its value at zero energy, $F_W(0)=1$. If a sizeable reduction in $F_W\lwig 0.02$ is realized
in nature, it is even possible to obtain a large cross-section of hadronic size, 
$\hat\sigma^{(I)}_{\rm ff}\gwig 1$~mb, which is observable in present day or near future 
cosmic ray facilities and neutrino telescopes in the form of cosmic proton or cosmic neutrino initiated 
events~\cite{Morris:1991bb,%Morris:1993wg
Fodor:2003bn}. 

This numerical fact is clearly demonstrated by the result for $\hat\sigma^{(I)}_{\rm ff}$ obtained in
Ref.~\cite{Ringwald:2002sw} (cf. Fig.~\ref{holy-grail} (bottom; solid and dotted)), 
where the tunneling suppression exponent has been estimated 
via the optical theorem from the $I\overline{I}$ interaction, known in pure $SU(N_c)$ 
gauge theory for arbitrary separation~\cite{Khoze:1991mx,Verbaarschot:1991sq%,Verbaarschot:1991sq,Balitsky:yz
}, 
(cf. Fig.~\ref{holy-grail} (top; solid)),   
and where the complete pre-exponential factor, with inclusion of its energy dependence, 
has been calculated in the saddle-point approximation. 
First of all, we see that the approximation of the pre-exponential factor displayed in Eq.~(\ref{cross-qfd}), 
leading to the estimate shown as a dotted line in Fig.~\ref{holy-grail} (bottom), gives in fact  
a reliable estimate of the complete pre-exponential factor over almost the full energy range.  
Moreover, the estimate of the cross-section becomes indeed of order $10^{-3}$~fb ($1$~mb)  
as soon as the estimate of the tunneling suppression exponent becomes of order $0.12$ ($0.02$).  

The analytical estimate based on the $I\overline{I}$ interaction in pure gauge theory is based on a
number of assumptions which, at energies above the sphaleron, $\epsilon\gwig 0.3$, may or may not be valid. 
Another method to infer the  tunneling suppression exponent was proposed in 
Refs.~\cite{Rubakov:1991fb,Tinyakov:1991fn,Rubakov:ec} and the results of a respective 
extended quantitative study was presented recently in 
Ref.~\cite{Bezrukov:2003er%,Bezrukov:2003qm
}. The method is based on the observation that the inclusive probability $P(\sqrt{\hat s},n)$ 
of tunneling from a state with energy $\sqrt{\hat s}$ and number of incoming particles
$n$ is calculable semi-classically, provided that $\epsilon \equiv \sqrt{\hat s}/(4\pi m_W/\alpha_W)$ 
and $\nu \equiv n\,\alpha_W$ are held fixed in the limit $\alpha_W\to 0$. In this regime, the 
probability $P(\sqrt{\hat s},n)$ has also an exponential form,
\begin{equation}
\label{incl-prob}
P(\sqrt{\hat s},n) \propto 
\exp\left[ -\frac{4\pi}{\alpha_W}\,
F_W \left( \epsilon , \nu \right)\right]
\,,
\end{equation}          
and the corresponding tunneling suppression exponent $F_W(\epsilon ,\nu)$ can be obtained 
by solving a classical boundary value problem for the Yang-Mills equations. 
Furthermore, it has been conjectured -- and proven in the first few orders of the perturbative 
expansion in fractional powers of $\epsilon$ 
(cf. Eq.~(\ref{FW-pert}))~\cite{Tinyakov:1991fn,Mueller:1992sc%,Bezrukov:2003zn
} as well as by comparison with the full quantum mechanical solution in a 
model with two degrees of freedom~\cite{Bonini:1999kj%,Bezrukov:2003yf
}
-- that the tunneling suppression exponent for the
two-particle cross-section is recovered in the limit of a small number of incoming
particles,
\begin{equation}
\label{F-lim}
F_W \left( \epsilon \right) = \lim_{\nu\to 0}
F_W \left( \epsilon , \nu \right)
\,.
\end{equation} 

In Ref.~\cite{Bezrukov:2003er%,Bezrukov:2003qm
}, the relevant classical boundary value problem  was solved, for spatially spherically  
symmetric\footnote{Without this simplification provided by spherical symmetry, the computational cost of the
numerical analysis seems to be prohibitive at present.}  
configurations of the $SU(2)$-Higgs gauge theory, for a large range of $\epsilon$ and $\nu$. 
Though computational limitations did not allow to reach literally the limit $\nu =0$, the authors
were able to extrapolate their results for the tunneling suppression exponent for multiple incoming
particles, $F_W \left( \epsilon , \nu \right)$, to zero $\nu$ and to get thereby a stringent 
lower bound on $F_W(\epsilon )$ (cf. Fig.~\ref{holy-grail} (top; dashed shaded)). In addition, they 
provided an independent estimate for $F_W(\epsilon )$ (cf. Fig.~\ref{holy-grail} (top; dashed)). 

It is important to note that the lower bound on and the estimate of $F_W(\epsilon )$ obtained in 
Ref.~\cite{Bezrukov:2003er%,Bezrukov:2003qm
} are applicable, strictly speaking, only for s-wave scattering, since they  
have been obtained exploiting spatially spherically symmetric configurations. 
This property is shared by the dominant intermediate saddle-point configurations
encountered in the $I\overline{I}$ ``valley'' 
configuration~\cite{Khoze:1991mx,Verbaarschot:1991sq%,Verbaarschot:1991sq,Balitsky:yz
}.
Therefore, the estimate of $F_W(\epsilon )$ 
from Ref.~\cite{Bezrukov:2003er%,Bezrukov:2003qm
} should, at small energies, coincide with the analytical estimate based on the $I\overline{I}$ interaction. 
This is clearly demonstrated in Fig.~\ref{holy-grail} (top; dashed and solid, respectively). 
At higher energies, above the sphaleron, $\epsilon \gwig 0.3$, this coincidence is no more observed. 
Apparently, the estimate based on the $I\overline{I}$ interaction, which has been argued 
in Ref.~\cite{Ringwald:2002sw} to be reliable for $\epsilon\lwig 0.75$, gives a 
more optimistic value for the two-particle tunneling suppression exponent than the one from the 
extrapolation of $F_W \left( \epsilon , \nu \right)$ to $\nu =0$.

%\section{An upper bound on the cross-section}
{\bf 4.}
At energies above the sphaleron, $\epsilon\gwig 0.3$, the total cross-section of electroweak
baryon number violation 
may well be dominated by axial symmetric configurations~\cite{Gould:1993hb}, 
on which there is at present no information available. 
Nevertheless, it is very instructive to exploit the lower bound on the tunneling 
suppression exponent for s-wave scattering from Ref.~\cite{Bezrukov:2003er%,Bezrukov:2003qm
} and to determine from it, via Eq.~(\ref{cross-qfd}), an upper bound on the cross-section of 
s-wave electroweak $B+L$ violating processes initiated by the scattering of two fermions. 
We find that the estimate based on the $I\overline{I}$ interaction violates this s-wave
upper bound for $\epsilon\gwig 0.9$ (cf. Fig.~\ref{holy-grail} (bottom)). 
However, as shown in Fig.~\ref{holy-grail} (bottom; dashed shaded), the upper bound is 
rapidly getting less stringent and exceeds $\gwig 10^{-3}$~fb 
for $\epsilon \simeq \sqrt{\hat s}/(30\ {\rm TeV})\gwig 1.2$. Therefore, it  
does not exclude the observability of electroweak $B+L$ violation at the VLHC. 
The upper bound even grows above $1$~mb ($100$~mb) for 
$\epsilon \gwig 7.5$ ($\epsilon \gwig 9.7$), thus still allowing for observable 
consequences of anomalous electroweak processes in ultrahigh energy 
($E = \hat s/(2\,m_p)\,\gwig\, 5\times 10^{19}$~eV) cosmic ray and 
neutrino physics~\cite{Morris:1991bb,%Morris:1993wg
Fodor:2003bn}.  
However, in order to reach such an ${\mathcal O}({\rm mb})$ cross-section, as also 
suggested by the $I\overline{I}$ estimate in Fig.~\ref{holy-grail} (bottom),
a substantial contribution from higher partial waves is 
required. The $I\overline{I}$ estimate and the upper bound violate the s-wave unitarity 
bound for the inelastic cross-section,
\begin{equation}
\label{unit-lim}
\hat\sigma^{(\ell =0)}_{\rm inel} < \frac{4\pi}{\hat s}
\simeq 5.5\ {\rm pb}\  \frac{1}{\epsilon^2} 
\,,
\end{equation}          
above $\epsilon \gwig 1.1$ and $\epsilon \gwig 1.9$, respectively.
    
\section*{Acknowledgements}
I would like to thank F.~Bezrukov, V.~V.~Khoze, V.~Rubakov, P.~Tinyakov, 
and F.~Schrempp for fruitful discussions, helpful comments, and a careful 
reading of the manuscript. I would also like to thank the authors of 
Ref.~\cite{Bezrukov:2003er%,Bezrukov:2003qm
} for providing their data concerning the tunneling suppression exponent.

\end{document}